# Beamforming in MISO Systems: Empirical Results and EVM-based Analysis

Melissa Duarte, Ashutosh Sabharwal, Chris Dick, and Raghu Rao

## Abstract

We present an analytical, simulation, and experimental-based study of beamforming Multiple Input Single Output (MISO) systems. We analyze the performance of beamforming MISO systems taking into account implementation complexity and effects of imperfect channel estimate, delayed feedback, real Radio Frequency (RF) hardware, and imperfect timing synchronization. Our results show that efficient implementation of codebook-based beamforming MISO systems with good performance is feasible in the presence of channel and implementation-induced imperfections. As part of our study we develop a framework for Average Error Vector Magnitude Squared (AEVMS)-based analysis of beamforming MISO systems which facilitates comparison of analytical, simulation, and experimental results on the same scale. In addition, AEVMS allows fair comparison of experimental results obtained from different wireless testbeds. We derive novel expressions for the AEVMS of beamforming MISO systems and show how the AEVMS relates to important system characteristics like the diversity gain, coding gain, and error floor.

This work of first two authors was partially supported by NSF Grants CNS-0551692 and CNS-0619767. The first author was also supported by a Xilinx Fellowship and a Roberto Rocca Fellowship. The authors also thank Azimuth Systems for providing the channel emulator used in this work.

M. Duarte and A. Sabharwal are with the Department of Electrical and Computer Engineering, Rice University, Houston, TX, 77005 USA, e-mail: {mduarte, ashu}@rice.edu.

C. Dick and R. Rao are with Xilinx Inc., San Jose, CA, 95124 USA, e-mail: {chris.dick, raghu.rao}@xilinx.com.

## I. Introduction

Standards for next generation wireless communications have considered the use beamforming Multiple Input Single Output (MISO) systems with codebook-based feedback because these systems can potentially achieve same diversity order and larger coding gain compared to non-feedback systems like space-time codes [1–4]. Recently, the performance of beamforming MISO systems has been analyzed taking into account errors in the channel estimate, and/or feedback delay [5–9], and noise in the feedback channel [10]. However, these results do not take into account effects of non-ideal RF processing, imperfect timing synchronization or consider implementation complexity.

In this paper we evaluate the performance of codebook based beamforming MISO systems taking into account implementation complexity and the presence of channel and implementation-induced imperfections. Specifically, we consider channel-induced imperfections which are due to channel estimation errors and feedback delay and we consider implementation-induced imperfections which are a result of imperfect timing synchronization and non-ideal RF processing, Automatic Gain Control (AGC), Analog to Digital Converters (ADCs), and Digital to Analog Converters (DACs). Since not all imperfections can be modeled tractably, especially implementation-induced imperfections, we adopt a mixed approach of analytical, simulation, and experimental evaluation. Analytical and simulation results presented in this paper take into account channel-induced imperfections but do not take into account implementation-induced imperfections because these imperfections are difficult to model in a tractable way. Thus, we complement these results with experimental results which do take into account both channel and implementation-induced imperfections. This mixed approach provides a more complete picture of expected performance.

Inclusion of experimental evaluation poses a unique challenge in the choice of evaluation metric. Common metrics like Bit Error Rate (BER) or Symbol Error Rate (SER) are usually analyzed as a function of the average Energy per Symbol to Noise ratio ($E_s/N_o$) or average Energy per Bit to Noise ratio ($E_b/N_0$). However, when real hardware is used for evaluation of wireless systems, getting an accurate measurement of the noise or the $E_s/N_o$ or $E_b/N_o$ proves problematic because the noise can be non-linear, both multiplicative and additive, and may depend on radio settings and characteristics of the received signal. In contrast, the Average Error

Vector Magnitude Squared (AEVMS), a metric commonly used in test equipment, can be easily measured since it is computed at the input of the demodulator. As a result, we propose to use the AEVMS as a metric for performance analysis. This leads to the natural question regarding the relationship between AEVMS and $E_s/N_o$ or $E_b/N_o$.

Our first contribution is a framework for AEVMS-based analysis of beamforming MISO systems. Although the Error Vector Magnitude (EVM) and EVM-based metrics are heavily used in industry for testing of wireless devices [11, 12], there is very little theory behind the use of EVM for performance analysis. Some previous work can be found in [11–16] but no previous work has analyzed the performance of beamforming MISO systems using an EVM related metric. We present simulation, analytical, and experimental results that show how the AEVMS relates to the $E_s/N_o$, BER, diversity gain, coding gain, and error floor. Since BER and AEVMS are quantities that can be directly measured, using these two metrics allows a straightforward comparison of analytical, simulation, and experimental results on the *same* scale. Furthermore, using metrics like BER and AEVMS facilitates comparison of results obtained with different wireless testbeds because these metrics are usually easy to measure in any testbed. We show that BER vs. (1/AEVMS) results can be used to analyze the diversity gain of a system. We also show that coding gain and error floors can be analyzed by looking at the AEVMS performance as a function of the $E_s/N_o$ or an $E_s/N_o$ related metric like the signal power.

Our second contribution is the performance analysis of beamforming MISO systems as a function of the amount of training used for channel estimation. In particular, we consider two different beamforming systems: a 1 round (1R) system which uses only 1 round of training and a 1.5 round (1.5R) system which uses 1.5 rounds of training (we use the terminology for multi round training defined in [17]). We present novel results on the AVEMS vs. $E_s/N_o$ performance of the 1R and 1.5R systems in the presence of channel estimation errors and feedback delay. These results show that in the presence of feedback delay, 1.5 rounds of training eliminate the error floor that is present when only one round of training is used. Taking into account noisy channel estimate and feedback delay, work in [5] analyzed the BER and SER for a 1R system and work in [8] analyzed the capacity of a 1.5R system. However, previous work does not include comparison and AEVMS-based analysis of error floor of 1R and 1.5R systems in the presence of imperfect channel estimate and feedback delay.

Our third contribution is an experimental evaluation which demonstrates that efficient imple-mentation of codebook based beamforming MISO systems with good performance is feasible in the presence of channel and implementation-induced imperfections. We show that beamforming codebooks proposed in [18, 19], which are known to facilitate efficient implementation and storage, can achieve performance close to infinite feedback (infinite codebook size) using only few feedback bits (small codebook size) and have better performance than a space-time code system like Alamouti. This result had not been demonstrated in the presence of channel and implementation-induced imperfections. Experimental results for beamforming systems have been reported in [20, 21] but these works have not considered codebook based feedback. We also consider the tradeoff between implementation complexity and performance in WiMAX compliant systems. Our experimental results demonstrate that the Mixed Codebook scheme for WiMAX compliant systems proposed in [19, 22] has good performance and simplifies implementation of beamforming in WiMAX compliant systems.

The rest of the paper is organized as follows. Section II describes the channel model and implementation requirements for the beamforming systems that are considered in this paper. The framework for AEVMS-based analysis of beamforming MISO systems is presented in Section III, this section also presents error floor analysis of 1R and 1.5R systems. Section IV describes the experimental setup and presents experiment results. Conclusions are presented in Section V.

## II. Beamforming System: Model and Codebooks

### A. Channel Model, Channel Estimation and Feedback Delay

We consider a MISO system with $T$ transmit antennas and one receive antenna. The received signal at time $k$ is equal to $r[k] = \mathbf{h}[k]\mathbf{x}[k] + n[k]$, where the $T \times 1$ vector $\mathbf{x}[k]$ represents the transmitted signal at time $k$, $\mathbf{h}[k]$ is the $1 \times T$ MISO channel at time $k$, and $n[k]$ represents the additive white Gaussian noise (AWGN) at the receiver, which is distributed as $n \sim \mathcal{CN}(0, N_o)$. The channel vector $\mathbf{h}[k]$ is given by $\mathbf{h}[k] = [h_1[k], h_2[k], ..., h_T[k]]$, where $h_i \sim \mathcal{CN}(0, \Omega)$ and the entries of $\mathbf{h}[k]$ are i.i.d. Thus, $\mathbf{h} \sim \mathcal{CN}(\mathbf{0}, \Omega\mathbf{I})$.

In this paper, we consider closed-loop beamforming based on receiver feedback. Using a unit norm $1 \times T$ beamforming vector $\mathbf{w}[k]$, the vector input to the channel is determined as $\mathbf{x}[k] = \sqrt{E_s}\mathbf{w}^\dagger[k]s[k]$, where $s[k]$ denotes the normalized constellation symbol transmitted at time $k$ ($E[|s[k]|^2] = 1$), $E_s$ is the average energy of the transmitted signal $\mathbf{x}[k]$ ($E[\|\mathbf{x}[k]\|^2] = E_s$), and $()^\dagger$ denotes matrix transpose. Beamforming vectors are part of a predetermined codebook, known to both the transmitter and receiver prior to communication. Furthermore, the codebook is considered to be fixed throughout the communication.

Since the channel is time-varying and unknown a priori, the receiver has to estimate the channel based on training signals. Using training signals sent orthogonal in time with energy $E_p$ and assuming AWGN at the receiver, the channel estimate at the receiver is given by

$$\widehat{\mathbf{h}}[k] = \mathbf{h}[k] + \boldsymbol{\Delta}\mathbf{h}[k], \tag{1}$$

where $\boldsymbol{\Delta}\mathbf{h}[k]$ represents the noise in the channel estimate distributed as $\boldsymbol{\Delta}\mathbf{h} \sim \mathcal{CN}(\mathbf{0}, \sigma_e^2\mathbf{I})$ and $\sigma_e^2$ depends on the training signal energy $E_p$ to noise energy $N_o$ ratio [23]. Thus, the channel estimate is distributed as $\widehat{\mathbf{h}} \sim \mathcal{CN}(\mathbf{0}, \Lambda\mathbf{I})$, with $\Lambda = \Omega + \sigma_e^2$. The channel estimate in (1) applies to both Minimum Mean Squared Error estimator and Maximum Likelihood estimator. In general, the training signal energy $E_p$ is not exactly equal to the signal energy $E_s$. For example, in WiMAX systems the training energy is 2.5 dB higher than $E_s$. Hence, we assume $E_p \propto E_s$ and

$$\sigma_e^2 \propto (E_s/N_o)^{-1}. \tag{2}$$

To account for errors in channel estimation and delay in the feedback channel, we use the model presented in [5]. In the presence of a feedback delay of $D$ seconds and noisy channel estimate as given in (1) we can write

$$\mathbf{h}[k] = \rho\sqrt{\frac{\Omega}{\Lambda}}\widehat{\mathbf{h}}[k-D] + \sqrt{(1-|\rho|^2)\Omega}\mathbf{v}[k-D], \tag{3}$$

where $\mathbf{v} \sim \mathcal{CN}(\mathbf{0}, \mathbf{I})$ and $\rho$ is the complex correlation coefficient given by $\rho = \frac{E[h_i[k]\widehat{h}_i[k-D]^*]}{\sqrt{\Omega\Lambda}}$ and we use $()^*$ to denote conjugate transpose. As was shown in [5], the correlation coefficient $\rho$ is related to the delay-only correlation coefficient $\rho_d$ and the estimation-error-only correlation coefficient $\rho_e$ as $\rho = \rho_d\rho_e$ where $\rho_d = \frac{E[h_i[k]h_i[k-D]^*]}{\Omega}$ and $\rho_e = \frac{E[h_i[k]\widehat{h}_i[k]^*]}{\sqrt{\Omega\Lambda}}$. Notice that $\rho_e$ can be written in terms of $\Omega$ and $\sigma_e^2$ as [24] $\rho_e = \sqrt{\Omega/(\Omega + \sigma_e^2)}$ and $\rho_d$ does not depend on $E_s/N_o$ but $\rho_e$ does. Using (2) we have that $\lim_{\frac{E_s}{N_o} \to \infty} \rho_e = 1$.

## B. Beamforming with Imprecise Information

We consider a beamforming MISO system with $B$ bits of feedback. The beamforming vector $\mathbf{w}[k]$ is chosen from a codebook of cardinality $N = 2^B$. We use an $N \times T$ matrix $\mathbf{W}$ to represent a codebook for a system with $T$ transmit antennas and codebook size $N$, and we use $\mathbf{w}_i$ to represent the $i$-th row of matrix $\mathbf{W}$. The beamforming codebook $\mathbf{W}$ is known to both the transmitter and the receiver. The channel estimate at the receiver at time $k - D$ is quantized into one of the codewords in the codebook, quantization is performed via an exhaustive search over the codewords in the codebook [3],

$$b = \arg \max_{1 \leq i \leq N} \left| \widehat{\mathbf{h}}[k - D] \mathbf{w}_i^\dagger \right|^2. \tag{4}$$

Index $b$ output by the channel quantizer is feedback to the transmitter and the transmitter chooses vector $\mathbf{w}_b$ for beamforming (we assume error-free feedback channel). Hence, with a feedback delay of $D$, the beamforming vector used at time $k$ is $\mathbf{w}[k] = \mathbf{w}_b$ and the received signal at time $k$ is equal to

$$r[k] = \mathbf{h}[k] \mathbf{w}_b^\dagger \sqrt{E_s} s[k] + n[k]. \tag{5}$$

In the case of infinite feedback ($N = \infty$) the beamforming vector is given by $\mathbf{w}[k] = \frac{\widehat{\mathbf{h}}[k-D]^*}{||\widehat{\mathbf{h}}[k-D]||}$.

## C. Codebooks

We consider four different types of beamforming codebooks which represent a tradeoff between performance and implementation complexity, and represent the best known methods spanning the two metrics. Specifically, we choose Maximum Welch Bound Equality (MWBE), WiMAX , Equal Gain Bipolar (EGB), and Tripolar codebooks. Description of these codebooks is shown in Table I. The EGB and Tripolar codebooks can be generated using the vector mapping techniques in [19]. Also, EGB codebooks can be designed via a Kerdock code construction [18].

Our previous work [19] showed that implementation of the channel quantization operation in (4) can require a large amount of complex multiplications depending on the codebook structure. Table I shows the amount of resources required for channel quantization for the four types of codebooks considered. Using an MWBE or a WiMAX codebook requires a large amount of complex multipliers. In contrast, using an EGB or a Tripolar codebook does not require any complex multipliers. EGB and Tripolar codebooks allow implementation of complex multiplications for channel quantization using simple multiplexers as was shown in [19]. The EGB codebook requires the least amount of resources among the four types of codebooks considered.

For the Rayleigh i.i.d channel described in Section II-A and assuming an ideal scenario where there are no channel or implementation-induced imperfections, it is known that all the codebooks considered in this paper have similar performance [2–4, 18, 19, 25–27]. WiMAX and MWBE codebooks result in slightly better performance than EGB or Tripolar codebooks, but the performance difference is usually less than 0.5 dB. Hence, in this ideal scenario, EGB codebooks are a good design choice because of their good performance and efficient implementation. In this paper we will investigate how MWBE, WiMAX, EGB, and Tripolar codebooks perform in the presence of channel and implementation-induced imperfections.

We will also present experimental evaluation of the WiMAX Mixed Codebook scheme proposed in our previous work in [19, 22]. In a WiMAX Mixed Codebook scheme, the WiMAX codebook is used at the transmitter for beamforming while channel quantization at the receiver is implemented using an EGB or a Tripolar codebook which is obtained by mapping the WiMAX codebook. The mapping is performed as proposed in [19]. The Mixed Codebook scheme remains WiMAX compliant because the mapped WiMAX codebook is only used for channel quantization. For more details on the WiMAX Mixed Codebook scheme please refer to [19, 22].

## III. Error Vector Magnitude Analysis of Beamforming MISO Systems

For a receiver demodulator using a normalized constellation, the AEVMS is given by

$$\text{AEVMS} = E[|s[k] - \widehat{s}[k]|^2], \tag{6}$$

where $\widehat{s}[k]$ is the decision variable which is input to the demodulator. In this section we analyze the AEVMS to understand the error floor, BER, diversity gain, and coding gain of a beamforming MISO system. The framework for AEVMS-based analysis developed in this section will be used for analysis of experimental results presented in Section IV.

### A. Training

The received signal $r[k]$ is equalized to obtain the decision variable $\widehat{s}[k]$ which is input to the demodulator. The value used for equalization depends on the channel estimate obtained from training. We consider a 1R system and a 1.5R system which differ in the amount of training that is used for channel estimation. The two systems are explained below.

*1) 1R system:* Fig. 1(a) shows a frame for a 1R system. The 1R system uses only one training sequence and the channel is estimated only once per frame. The channel estimate is computed at the receiver upon reception of the training sequence. The channel estimate is used for computation of the feedback information and for equalization. The training sequence consists of transmission of a training signal from each transmitter antenna and training signals from different antennas are sent orthogonal in time. Two preambles are transmitted, one before the training sequence and one before the payload, the preambles are used for AGC and timing synchronization. The second preamble and the payload are beamformed.

*2) 1.5R system:* This system uses two training sequences, as shown in Fig. 1(b). The first training sequence is exactly the same as the training sequence used in the 1R system. The second training sequence is the beamformed version of the first training sequence. Two preambles, used for AGC and timing synchronization, are transmitted before each training sequence. The second preamble and the payload are beamformed. In the 1.5R system the channel is estimated twice. The first estimate is computed from the first training sequence and this estimate is used for computation of the feedback information. The second channel estimate is computed after the second training sequence, since the second training sequence is beamformed, the second channel estimate corresponds to the equivalent beamformed channel. The estimate of the equivalent beamformed channel is used for equalization.

The second training sequence is beamformed because of two reasons. First, beamforming the second training sequence simplifies the implementation of the equalizer. Second, although we assume noiseless feedback, estimating the equivalent beamformed channel may be useful in a system with noisy feedback where the codeword chosen by the receiver is not the codeword being used by the transmitter. In a noisy feedback system it may be better to get an estimate of the equivalent beamformed channel and use this channel estimate for equalization, instead of estimating the channel and then computing the equalization signal using this estimate and the codeword chosen by the receiver. This intuition is based on results presented in [17] for feedback based power control schemes, where it is shown that using power controlled training improves performance in a noisy feedback system.

We use $\mathbf{q}[k]$ to denote the beamformed channel at time $k$. The $1 \times T$ vector $\mathbf{q}[k]$ is given by $\mathbf{q}[k] = [q_1[k], q_2[k], ..., q_T[k]]$ where $q_i[k] = h_i[k]w_{b,i}$ and $w_{b,i}$ denotes the $i$-th entry of vector $\mathbf{w}_b$. The additive Gaussian noise in the estimation of $\mathbf{q}[k]$ is same as in the estimation of $\mathbf{h}[k]$ and the estimator used is also the same. Hence, the estimate of the equivalent beamformed channel is given by $\hat{\mathbf{q}}[k] = \mathbf{q}[k] + \boldsymbol{\Delta}\mathbf{q}[k]$, where $\boldsymbol{\Delta}\mathbf{q} \sim \mathcal{CN}(\mathbf{0}, \sigma_e^2 \mathbf{I})$.

### B. Equalization and Decision Variable

We now compute the decision variable for the 1R system and for the 1.5R system. Results presented in this section and in Sections III-C to III-E take into account channel-iduced imperfections and assume there are no implementation-induced imperfections.

*1) 1R system:* The receiver knows $\widehat{\mathbf{h}}[k-D]\mathbf{w}_b^{\dagger}\sqrt{E_s}$ and the decision variable is equal to $\widehat{s}_{1R}[k] = r[k]\frac{\left(\widehat{\mathbf{h}}[k-D]\mathbf{w}_b^{\dagger}\sqrt{E_s}\right)^*}{\left|\widehat{\mathbf{h}}[k-D]\mathbf{w}_b^{\dagger}\sqrt{E_s}\right|^2}$. Substituting $r[k]$ and $\mathbf{h}[k]$ using (5) and (3) respectively we obtain.

$$
\begin{aligned}
\widehat{s}_{1R}[k] &= \rho\sqrt{\frac{\Omega}{\Lambda}}s[k] + \sqrt{(1-|\rho|^2)\Omega}\,\mathbf{v}[k-D]\mathbf{w}_b^{\dagger}\frac{\left(\widehat{\mathbf{h}}[k-D]\mathbf{w}_b^{\dagger}\right)^*}{\left|\widehat{\mathbf{h}}[k-D]\mathbf{w}_b^{\dagger}\right|^2}s[k] \\
&\quad + \frac{n[k]}{\sqrt{E_s}}\frac{\left(\widehat{\mathbf{h}}[k-D]\mathbf{w}_b^{\dagger}\right)^*}{\left|\widehat{\mathbf{h}}[k-D]\mathbf{w}_b^{\dagger}\right|^2},
\end{aligned}
\tag{7}
$$

*2) 1.5R system:* The receiver has knowledge of $\widehat{a}[k]\sqrt{E_s}$ where $\widehat{a}[k] = \sum_{i=1}^{T}\widehat{q}_i[k] = \sum_{i=1}^{T}q_i[k] + \sum_{i=1}^{T}\Delta q_i[k] = a[k] + \Delta a[k]$. We use $\widehat{q}_i[k]$ and $\Delta q_i[k]$ to denote the $i$-th entry of vectors $\widehat{\mathbf{q}}[k]$ and $\boldsymbol{\Delta}\mathbf{q}[k]$ respectively and we define $a[k] = \sum_{i=1}^{T}q_i[k]$ and $\Delta a[k] = \sum_{i=1}^{T}\Delta q_i[k]$. Using (5) and the expressions for $a[k]$ and $\widehat{a}[k]$ above, we obtain the decision variable for the 1.5R system

$$
\widehat{s}_{1.5R}[k] = r[k]\frac{(\widehat{a}[k]\sqrt{E_s})^*}{|\widehat{a}[k]\sqrt{E_s}|^2} = \frac{a[k]}{\widehat{a}[k]}s[k] + \frac{1}{\widehat{a}[k]\sqrt{E_s}}n[k].
\tag{8}
$$

### C. AEVMS of a beamforming MISO System

In this section we compute the AEVMS for the 1R and 1.5R systems.

*1) 1R system:* Using (6) with $\widehat{s}[k]$ substituted with (7) we obtain

$$
\begin{aligned}
\text{AEVMS}_{1R} &= E\Bigg[\bigg|\bigg(1 - \rho\sqrt{\frac{\Omega}{\Lambda}} - \sqrt{(1-|\rho|^2)\Omega}\,\mathbf{v}[k-D]\mathbf{w}_b^{\dagger}\frac{\left(\widehat{\mathbf{h}}[k-D]\mathbf{w}_b^{\dagger}\right)^*}{\left|\widehat{\mathbf{h}}[k-D]\mathbf{w}_b^{\dagger}\right|^2}\bigg)s[k] \\
&\quad - \frac{n[k]}{\sqrt{E_s}}\frac{\left(\widehat{\mathbf{h}}[k-D]\mathbf{w}_b^{\dagger}\right)^*}{\left|\widehat{\mathbf{h}}[k-D]\mathbf{w}_b^{\dagger}\right|^2}\bigg|^2\Bigg].
\end{aligned}
\tag{9}
$$

Since $s[k]$, $n[k]$, $\mathbf{v}[k-D]$, and $\widehat{\mathbf{h}}[k-D]$ are independent and $E[|s[k]|^2] = 1$, $E[|n[k]|^2] = N_o$, and $E[n[k]] = 0$, we can simplify the expression above to obtain

$$
\begin{aligned}
\text{AEVMS}_{1R} &= 1 - 2\text{Re}\{\rho\}\sqrt{\frac{\Omega}{\Lambda}} + |\rho|^2\frac{\Omega}{\Lambda} + (1-|\rho|^2)\Omega E\left[\frac{\left|\mathbf{v}[k-D]\mathbf{w}_b^{\dagger}\right|^2}{\left|\widehat{\mathbf{h}}[k-D]\mathbf{w}_b^{\dagger}\right|^2}\right] \\
&\quad + E\left[\frac{1}{\left|\widehat{\mathbf{h}}[k-D]\mathbf{w}_b^{\dagger}\right|^2}\right]\frac{N_o}{E_s}.
\end{aligned}
\tag{10}
$$

In the Appendix we show that

$$E\left[\frac{\left|\mathbf{v}[k-D]\mathbf{w}_b^\dagger\right|^2}{\left|\widehat{\mathbf{h}}[k-D]\mathbf{w}_b^\dagger\right|^2}\right] = E\left[\frac{1}{\left|\widehat{\mathbf{h}}[k-D]\mathbf{w}_b^\dagger\right|^2}\right] \tag{11}$$

and

$$E\left[\frac{1}{\left|\widehat{\mathbf{h}}[k-D]\mathbf{w}_b^\dagger\right|^2}\right] = \frac{1}{\Lambda(T-1)}\Psi \tag{12}$$

where

$$\Psi = \left(1 + \frac{T-1}{T}N^{\frac{-1}{T-1}} \; {}_2F_1\left(1, T, 1+T, \left(\frac{1}{N}\right)^{\frac{1}{T-1}}\right)\right). \tag{13}$$

Substituting (11) and (12) in (10) we obtain $\text{AEVMS}_{1R}$ in closed form

$$\text{AEVMS}_{1R} = 1 - 2\text{Re}\{\rho\}\sqrt{\frac{\Omega}{\Lambda}} + |\rho|^2\frac{\Omega}{\Lambda} + (1-|\rho|^2)\Omega\frac{1}{\Lambda(T-1)}\Psi + \frac{1}{\Lambda(T-1)}\Psi\frac{1}{E_s/N_o}. \tag{14}$$

In the case of infinite feedback we have that $N = \infty$ hence $\Psi = 1$.

*2) 1.5R system:* For a 1.5R system the AEVMS, computed using (6) and (8), is equal to

$$\text{AEVMS}_{1.5R} = E\left[\left|1 - \frac{a[k]}{\widehat{a}[k]}\right|^2\right] + E\left[\frac{1}{|\widehat{a}[k]|^2}\right]\frac{1}{E_s/N_o}. \tag{15}$$

We do not have a closed form expression for $\text{AEVMS}_{1.5R}$ because we do not have a closed form expression for the expectations in (15). Computing these expectations is complicated because they depend on the channel estimate at time $k$ and the quantized channel estimate at time $k - D$. However, (15) allows us to do an asymptotic (large $E_s/N_o$) analysis of a 1.5R system and we also provide simulation results and experiment results that show how this system performs.

### D. Relation Between AEVMS and Error Floor

At infinite $E_s/N_o$ we expect the AEVMS to be equal to zero. If this is not the case then the system has an error floor. If the system has an error floor then it will not be possible to decrease the AEVMS below a certain value greater than zero no matter how large the $E_s/N_o$. We next show that in the ideal case of no feedback delay and no channel estimation error, the 1R and 1.5R systems do not have an error floor. However, when feedback delay and channel estimation errors are taken into account, the 1R system has an error floor while the 1.5R system does not.

*Definition 1:* An error floor exists if $\lim_{\frac{E_s}{N_o}\to\infty}\text{AEVMS} > 0$. An error floor does not exist if $\lim_{\frac{E_s}{N_o}\to\infty}\text{AEVMS} = 0$.

*Proposition 1:* For the 1R system, the following is true.

*(a)* In the ideal case of no feedback delay ($D = 0$ hence $\rho_d = 1$) and no channel estimation error ($\sigma_e^2 = 0$), the 1R system does not have an error floor and the AEVMS given by

$$\text{AEVMS}_{1R|\sigma_e^2=0,\rho_d=1} = \frac{1}{\Omega(T-1)}\Psi\frac{1}{E_s/N_o}. \tag{16}$$

*(b)* In the case of feedback delay ($|\rho_d| < 1$) and channel estimation error such that $\sigma_e^2$ satisfies (2), the 1R system has an error floor.

*Proof:* Part (a) follows from Definition 1 and simplification of (14) using $\rho_d = 1$ and $\sigma_e^2 = 0$. Part (b) follows from Definintion 1 and use of (14), $|\rho_d| < 1$, $\lim_{\frac{E_s}{N_o}\to\infty}\rho_e = 1$, and $\lim_{\frac{E_s}{N_o}\to\infty}\Lambda = \Omega$ to compute $\lim_{\frac{E_s}{N_o}\to\infty}\text{AEVMS}_{1R|\sigma_e^2\propto(E_s/N_o)^{-1},|\rho_d|<1} = |\rho_d - 1|^2 + (1-|\rho_d|^2)\frac{1}{(T-1)}\Psi > 0.$ ∎

*Proposition 2:* For the 1.5R system, the following is true.

*(a)* In the ideal case of no feedback delay ($D = 0$ hence $\rho_d = 1$) and no channel estimation error ($\sigma_e^2 = 0$), the 1.5R system does not have an error floor and the AEVMS given by

$$\text{AEVMS}_{1.5R|\sigma_e^2=0,\rho_d=1} = \frac{1}{\Omega(T-1)}\Psi\frac{1}{E_s/N_o}. \tag{17}$$

*(b)* In the case of feedback delay ($|\rho_d| < 1$) and channel estimation error such that $\sigma_e^2$ satisfies (2), the 1.5R system does not have an error floor.

*(c)* In the case of feedback delay ($|\rho_d| < 1$) and channel estimation error such that $\sigma_e^2$ satisfies (2), the AEVMS of the 1.5R system at high $E_s/N_o$ can be approximated as

$$\text{AEVMS}_{1.5R|\sigma_e^2\propto(E_s/N_o)^{-1},|\rho_d|<1} \approx E\left[\frac{1}{|a[k]|^2}\right]\frac{1}{E_s/N_o}. \tag{18}$$

*Proof:* Part (a) follows from Definition 1 and simplification of (15) using $\rho_d = 1$ and $\sigma_e^2 = 0$. Part (b) follows from Definition 1 and use of (15), $|\rho_d| < 1$, $\lim_{\frac{E_s}{N_o}\to\infty}\hat{a}[k] = a[k]$ and $\lim_{\frac{E_s}{N_o}\to\infty}E[1/|\hat{a}[k]|^2] = E[1/|a[k]|^2]$ to compute $\lim_{\frac{E_s}{N_o}\to\infty}\text{AEVMS}_{1.5R|\sigma_e^2\propto(E_s/N_o)^{-1},|\rho_d|<1} = 0$. Proof of part (c) is as follows. At high $E_s/N_o$ we can approximate $\sigma_e^2 \approx 0$, $E[|1-a[k]/\hat{a}[k]|^2] \approx 0$, and $E[1/|\hat{a}[k]|^2] \approx E[1/|a[k]|^2]$. Using these approximations in (15) we obtain (18). ∎

Proposition 1(b) and Proposition 2(b) are verified via simulation results shown in Fig. 2(a) and Fig. 2(b). Fig. 2 shows results for five different scenarios labeled as SN1, SN2, ... , SN5. Fig. 2(d) specifies the legend for Fig. 2(a), Fig. 2(b), and Fig. 2(c). All results in Fig. 2 correspond to 16-QAM modulation and $\Omega = 1$. The legend in Fig. 2(d) specifies if results were obtained via simulation or analytically and if results correspond to a 1R or a 1.5R system. Results shown in Fig. 2(a) and Fig. 2(b) show that for $\rho_d = 0.99$ and $\sigma_e^2 = (E_s/N_o)^{-1}/2$ the 1R system has an error floor and the 1.5R system does not have an error floor. Error floors are usually identified by observing the BER vs. $E_s/N_o$ performance. Equivalently, errors floors and their causes can be identified by analyzing the performance of the AEVMS as a function of the $E_s/N_o$.

Intuitively, results in Proposition 1(b) and Proposition 2(b) can be explained as follows. For equalization, the 1R system uses $\hat{\mathbf{h}}[k - D]\mathbf{w}_b^\dagger \sqrt{E_s}$ and the 1.5R system uses $\hat{a}[k] \sqrt{E_s}$. Ideally, $\mathbf{h}[k]\mathbf{w}_b^\dagger \sqrt{E_s}$ should be used for equalization. Since $\sigma_e^2 \to 0$ as $E_s/N_o \to \infty$, then for high $E_s/N_o$ the signal used for equalization in the 1.5R system approximates to the ideal value, while for the 1R system the signal used for equalization approximates to $\mathbf{h}[k - D]\mathbf{w}_b^\dagger \sqrt{E_s}$.

Results asymptotic in $E_s/N_o$ for the AEVMS of the 1.5R system in Proposition 2(c) are verified in Fig. 2(b). Notice that at high $E_s/N_o$ the AEVMS of the 1.5R system decays proportionally with the increase in $E_s/N_o$, as stated in (18). Fig. 2 also shows that analytical and simulation results for the 1R system match accurately. Analytical results presented in Fig. 2(a) were computed using the equations for BER derived in [5], these BER equations were also used to obtain the analytical results in Fig. 2(c) by writing $E_s/N_o$ in terms of the AEVMS using (14). Analytical results in results in Fig. 2(c) were computed using (14). In Fig. 2 we have only included analytical results for SN1 in order to avoid cluttering the graphs. We note that analytical results for finite $N$ were obtained assuming an optimum codebook (see Appendix) and we have verified that this assumption yields analytical results which match accurately simulation results for MWBE, WiMAX, EGB and Tripolar codebooks and different values of $N$, $T$, $\sigma_e^2$, $\rho_d$ and $\Omega$.

### E. Relation Between AEVMS, BER, Diversity Gain and Coding Gain

Based on preceding analysis, the BER vs. $E_s/N_o$ plot can be decomposed into two plots: BER vs. (1/AEVMS) and AEVMS vs. $E_s/N_o$. This decomposition is shown in Fig. 2. For example, from Fig. 2(a) we can observe that for SN4 a BER of $10^{-3}$ is obtained at $E_s/N_o \approx 20$ dB. Equivalently, one can first observe from Fig. 2(c) that for SN4 a BER of $10^{-3}$ is obtained at AEVMS $\approx -18$ dB (i.e an 1/AEVMS value of 18 dB) and then observe from Fig. 2(b) that an AEVMS of $-18$ dB is obtained at $E_s/N_o \approx 20$ dB. In order to decompose the BER vs. $E_s/N_o$ plot into BER vs. (1/AEVMS) and AEVMS vs. $E_s/N_o$, one computes (analytically) or measures (in simulation) the BER and AEVMS corresponding to a given $E_s/N_o$ value.

A system has a diversity gain $G_d$ and coding gain $G_c$ if at high $E_s/N_o$ the BER scales as [28]

$$\text{BER} \approx \Big( G_c \frac{E_s}{N_o} \Big)^{-G_d}.$$ (19)

Hence, when (19) is a valid approximation, the BER vs. $E_s/N_o$ plot captures the coding and diversity gain of a system and in a log-log scale this plot is well approximated by a straight line that decays with slope $G_d$ and has a horizontal shift of $G_c$ dB relative to the benchmark curve of $(E_s/N_o)^{-G_d}$ [29]. In this section we show that when (19) is a valid approximation and the AEVMS can be approximated as

$$\text{AEVMS} \approx \beta \frac{1}{E_s/N_o},$$ (20)

where $\beta$ is a positive and finite constant, we have the following. When the BER vs. $E_s/N_o$ plot is decomposed into BER vs. (1/AEVMS) and AEVMS vs. $E_s/N_o$ plots, we have that the diversity gain and the part of the coding gain that depends on the modulation scheme (constellation) are captured by the BER vs. (1/AEVMS) plot and the part of the coding gain that does not depend of the modulation scheme is captured by the AEVMS vs. $E_s/N_o$ plot. The following result shows the relation between AEVMS and diversity gain.

*Proposition 3:* If at high $E_s/N_o$ the AEVMS can be approximated as in (20) and the BER can be approximated as in (19), then at high $E_s/N_o$, or equivalently at high values of 1/AEVMS, the BER as a function of the AEVMS is given by

$$\text{BER} \approx \Big( G_c \beta \frac{1}{\text{AEVMS}} \Big)^{-G_d}$$ (21)

and the the diversity gain $G_d$ can be computed as

$$G_d = - \lim_{\frac{1}{\text{AEVMS}} \to \infty} \frac{\log \text{BER}}{\log \frac{1}{\text{AEVMS}}}.$$ (22)

Consequently, the BER vs. (1/AEVMS) curve plotted in a log-log scale decays with slope $G_d$ and the BER vs. (1/AEVMS) plot captures the diversity gain.

*Proof:* Solving for $E_s/N_o$ from (20) and substituting the result in (19) we obtain (21). (22) can be readily verified by substituting (21) in (22) and computing the limit. ∎

As an example, consider the case of no feedback delay and no channel estimation errors. In this case the SER (hence the BER) can be approximated as in (19) at high $E_s/N_o$ [29] and the 1R and 1.5R systems have the same AEVMS which satisfies (20) for all values of $E_s/N_o$ (as can be seen from (16) and (17)). Hence, by solving for $E_s/N_o$ from (16) or (17) and substituting in (19) we obtain that at high 1/AEVMS the BER for $\rho_d = 1$ and $\sigma_e^2 = 0$ can be approximated as

$$\text{BER} \approx \left( G_c \frac{1}{\Omega(T-1)} \Psi \frac{1}{\text{AEVMS}} \right)^{-G_d}. \tag{23}$$

In the presence of feedback delay and channel estimation errors, the AEVMS of the 1R system cannot be approximated as in (20) because the system has an error floor, as was shown in Propostion 1(b). Because of the error floor the BER of the 1R system does not decay to zero, as can be seen Fig. 2(a) and Fig. 2(c), and (19) is not a valid approximation. For the 1.5R system and taking into account feedback delay and channel estimation errors, Proposition 2(c) shows that the AEVMS at high $E_s/N_o$ can be approximated as in (20) with $\beta = E[1/|a[k]|^2]$. The BER for the 1.5R system taking into account feedback delay and noisy channel estimate is not known. However, we note that it was shown in [6] that for a 1.5R system with feedback delay and perfect channel estimation (19) is a valid approximation and the diversity gain reduces to one when $|\rho_d| < 1$. This is consistent with results in Fig. 2(a) which suggest that there is a loss in diversity gain (diversity gain less than $T$) as can be observed from the slope of the curves for the 1.5R system at high $E_s/N_o$. Notice that for the 1.5R system, the slope of decay of the BER vs. $E_s/N_o$ curves is the same as for the BER vs. (1/AEVMS) curves, hence, the diversity gain is captured by the BER vs. (1/AEVMS) curves, this is consistent with Proposition 3.

We now analyze the relation between AEVMS and coding gain. Part of the coding gain of a beamforming MISO system depends on the modulation (e.g. MPSK or MQAM) being used. We use $G_{c,m}$ to label the part of the coding gain that depends on the modulation being used and we use $G_{c,p}$ to label the part of the coding gain that does not depend on the modulation being used. $G_{c,p}$ is mainly due to power gain or effective increase in received signal power as defined in [28]. The following conjecture relates the AEVMS and the coding gain.

*Conjecture 1:* For a beamforming MISO system with channel estimation errors and feedback delay the following is true. If at high $E_s/N_o$ the AEVMS can be approximated as in (20) and the BER can be approximated as in (19), then at high $E_s/N_o$, or equivalently at high values of 1/AEVMS, $G_{c,m}$ and $G_{c,p}$ satisfy the following.

*(a)* $G_{c,m}$, which is the part of the coding gain that depends on the modulation being used, is captured by the BER vs. (1/AEVMS) curve.

*(b)* $G_{c,p}$, which is the part of the coding gain that does not depend on the modulation being used, is captured by the AEVMS vs. $E_s/N_o$ curve.

Results in (14) and (15) do show that the AEVMS vs. $E_s/N_o$ curve is independent of the modulation. A formal proof for Conjecture 1 would require a general expression for the coding gain of the 1R and 1.5R systems taking into account channel estimation errors and feedback delay, this general expression is not known. We have performed extensive simulations that indicate that Conjecture 1 will most likely hold. Below we give an example.

As can be observed from Fig. 2(a), at high $E_s/N_o$ the difference in performance between SN4 and SN5 is due to coding gain (at high $E_s/N_o$ the BER vs. $E_s/N_o$ curves differ only by a horizontal shift). Observe from Fig. 2(c) that the BER vs. (1/AEVMS) plots for SN4 and SN5 lie on top of each other. Hence, from Proposition 3 we have that SN4 and SN5 have the same $G_d$ and from Conjecture 1(a) we have that SN4 and SN5 have the same $G_{c,m}$, which is consistent with the fact that results for SN4 and SN5 are both for 16 QAM. Since $G_d$ and $G_{c,m}$ for SN4 and SN5 are the same, then the difference in coding gain between SN4 and SN5 is only due to a difference in $G_{c,p}$. Hence, from Conjecture 1(b), the horizontal shift between the BER vs. $E_s/N_o$ plots for SN4 and SN5 must be equal to the horizontal shift between the AEVMS vs. $E_s/N_o$ plots for SN4 and SN5 and this can be verified from Fig. 2(a) and Fig. 2(b).

## IV. Experiment Setup and Results

In Section III we presented an AEVMS-based analysis of beamforming MISO systems which accounted for effects of delay and channel estimation errors. In order to simplify analysis, we did not take into account implementation-induced imperfections. In this section we present an empirical evaluation of beamforming MISO systems which was conducted using WARP [30] and a wireless channel emulator [31, 32]. Using a channel emulator allowed us to control channel related parameters like $\Omega$ and $\rho_d$. In addition, using real hardware for transmission and reception of RF signals allowed us to obtain results which account for real-world hardware effects. Hence, our experimental results take into account both channel and implementation-induced imperfections. We present experiment results using the AEVMS-based framework we presented in Section III.

*A. Experiment Setup and Scenarios Considered*

Experiments were implemented using the WARPLab framework [33] which allows rapid prototyping of physical layer algorithms by combining the ease of MATLAB with the capabilities of WARP. The WARPLab framework provides the software necessary for easy interaction with the WARP nodes directly from the MATLAB workspace, the software consists on FPGA code and MATLAB m-code functions, which are all available in the WARP repository [34]. Two WARP nodes were used, one as a transmitter node and the other one as the receiver node. The main component of the WARP node hardware is a Xilinx Virtex-II Pro FPGA. Each node also has four daughter card slots, each slot is connected to a dedicated bank of I/O pins on the FPGA, these daughter card slots were used to connect the FPGA to up to four different radio boards. For our experiments, we used two and four radios at the transmitter to build a $2 \times 1$ and a $4 \times 1$ MISO system respectively. At the receiver, only one radio board was used.

The experiments were implemented using the basic WARPLab setup [33] where two WARP nodes are connected to a host PC via an Ethernet switch. The baseband waveforms (samples) were constructed in MATLAB and the samples were stored in buffers on the FPGA on the transmitter node, download of samples from the MATLAB workspace to the FPGA buffers was done using the software provided in the WARPLab framework. A trigger signal sent from the host PC to the WARP nodes started transmission of samples from the transmitter node and storage of received samples on buffers on the receiver node. The radio boards at the transmitter node upconverted the baseband samples to RF waveforms and the radios at the receiver node downconverted the received RF signal to baseband samples that were stored on the buffers on the receiver FPGA. The samples in the receive buffers were loaded to the MATLAB workspace on the host PC using functions from the WARPLab framework. Processing of the received baseband samples was done in MATLAB. The error-free feedback channel was implemented in the host PC.

Experiments were performed for a $2 \times 1$ and a $4 \times 1$ MISO system. We only implemented the 1.5R system in Fig. 1(b). The 1R system was not considered for experimental evaluation because, as was shown in Section III, the 1.5R system outperforms the 1R system for a large range of $E_s/N_o$ values. In order to compare the performance of a feedback-based system like beamforming with a non-feedback-based system like Alamouti, we also implemented and tested a $2 \times 1$ Alamouti scheme [35] using the WARPLab framework. For the Alamouti implementation only one training sequence was sent and payload was sent immediately after the training sequence was transmitted. The rest of the experiment conditions in the Alamouti implementation were equal to the experiment conditions in the beamforming implementation. Experiment conditions are shown in Table II. We note that the number of payload symbols per frame was limited to 110 due to the characteristics of the transmitted signal (128 samples per symbol plus samples used for training and preamble) and the maximum number of samples that can be stored per receiver radio in a WARP node ($2^{14}$ samples). The clock was shared between the transmitter and the receiver to avoid carrier frequency offset effects. The wireless channel emulator was set so that an RF link was enabled from each transmitter radio to the receiver radio in order to emulate a MISO system and each RF link consisted of three paths. Since the delay spread was much smaller than the symbol period the transmitted signal went through a flat fading channel. The emulated channel corresponds to the channel model described in Section II-A.

### B. Empirical Results Using a Wireless Channel Emulator

Results obtained using the channel emulator are presented in Fig. 3. Fig. 3(d) specifies the legend for Fig. 3(a), Fig. 3(b) and Fig. 3(c), the seven different scenarios that were evaluated via experiments are labeled as EXP1, EXP2, ..., EXP7. Fig. 3(c) also includes simulation results for a $2 \times 1$ and a $4 \times 1$ MISO system with $\sigma_e^2 = (E_s/N_o)^{-1}/2.3$ (noise variance of the channel estimate is 3.6 dB lower than $(E_s/N_o)^{-1}$ to match the fact that in experiments the total training signal energy per antenna was 3.6 dB larger than the total energy per symbol), $\rho_d = 0.9996$ (as in experiments), and $\Omega = 1$. Including the effect of the channel, the average energy per symbol is equal to $\Omega E_s$ and the average energy per symbol to noise ratio is equal to $\Omega E_s/N_o$. In the experiments, the emulator output power is equal to $\Omega E_s$.

Results in Fig. 3(a) verify that a feedback system like beamforming has better performance than a non feedback system like Alamouti. From results in Fig. 3(a) we also observe that the performance of MWBE, EGB, and WiMAX codebooks is approximately the same. For $T = 2$ and $N = 4$, the curves corresponding to the EGB codebook and the MWBE codebook are approximately the same, for parts of the curves it may seem that one codebook has better performance than the other but the curves cross each other several times indicating the performance for the two codebooks is approximately the same. Similarly, we observe that for $T = 4$ and $N = 64$ the EGB and the WiMAX codebook have similar performance.

Results for $T = 4$ in Fig. 3(a) show that the diversity gain with infinite feedback is the same as the diversity gain with finite feedback, since the BER curves appear to decay with the same slope (EXP4, EXP5, EXP6 and EXP7 decay with approximately same slope). The only difference between infinite and finite feedback is the coding gain, as can be observed from the horizontal shift for $T = 4$ curves in Fig. 3 (a) (shift of EXP7 with respect to EXP4, EXP5 and EXP6). The difference in performance between infinite feedback and finite feedback is between 1 dB and 2 dB for most of the average received signal powers considered.

Results in Fig. 3(a) show that the WiMAX Mixed Codebook scheme (EXP4) has worse performance than the WiMAX scheme (EXP5) and the performance loss is approximately 1dB. In the WiMAX Mixed Codebook scheme used to obtain EXP4 result, a Tripolar codebook was used for channel quantization. Using an EGB instead of a Tripolar codebook would allow a more efficient implementation but results in [19] showed that this would result in a worse performance of the Mixed Codebook scheme. There is a tradeoff between implementation complexity and performance; using a WiMAX Mixed Codebook scheme simplifies the implementation of the channel quantizer but results in a small performance degradation.

In the presence of channel and implementation-induced imperfections, results in Fig. 3(a) demonstrate that EGB codebooks have good performance. Since EGB codebooks also allow efficient implementation, we conclude that EGB codebooks are the best option out of the four types of codebooks considered. Results in Fig. 3(a) also demonstrate that in a WiMAX compliant system, a WiMAX Mixed Codebook scheme using a Tripolar codebook offers a good tradeoff between implementation complexity and performance.

BER vs. $\Omega E_s$ results in Fig. 3(a) can be used to compare different experimental results but are not useful to compare experimental results with simulation or analytical results. Translating $\Omega E_s$ values to $E_s/N_o$ or vice versa is complicated because measuring $N_o$ or $E_s/N_o$ is not straightforward since the noise can be non-linear, both multiplicative and additive and may depend on radio settings and characteristics of the received signal. Hence, translating the BER vs. $\Omega E_s$ results into BER vs. $E_s/N_o$ or vice versa proves problematic. To facilitate comparison between simulation and experimental results we decompose results in Fig. 3(a) into BER vs. (1/AEVMS) and AEVMS vs. $\Omega E_s$, as show in Fig. 3(c) and Fig. 3(b) respectively. In order to do this decomposition one measures the BER and the AEVMS for a given $\Omega E_s$. This decomposition is analogous to the one done in Section III-E where the BER vs. $E_s/N_o$ plot was decomposed into BER vs. (1/AEVMS) and AEVSM vs. $E_s/N_o$. BER and AEVMS are metrics that can be easily measured (the AEVMS is computed before the demodulator and the BER is computed after the demodulator) and are commonly measured in testing of wireless devices [11, 12]. Hence, using BER vs. (1/AEVMS) for performance analysis allows a straightforward comparison of experimental results with simulation results on the *same* scale, as shown in Fig. 3(c). Results in this figure show that experimental results match closely simulation results, there are some differences but these may be due to hardware effects that were not considered in simulations.

It is important to keep in mind that, as shown in Section III-E, part of the coding gain of a system is not captured by the BER vs. 1/AEVMS plots. Differences in coding gain that are not capture in the BER vs. 1/AEVMS plots can observed by plotting the AEVMS as a function of the $E_s/N_o$, as shown in Section III-E, or an $E_s/N_o$ related metric like $\Omega E_s$. As an example, consider results in Fig. 3(a) for $T = 4$. All curves for $T = 4$ decay with approximately the same slope and the main difference between curves is a horizontal shift, hence, the main difference between results is due to a difference in coding gain. However, curves for $T = 4$ in Fig. 3(c) are approximately the same, consequently at least part of the coding gain is not being captured by the BER vs. (1/AEVMS) plots. As can be seen in Fig. 3(b), part of the coding gain that is not captured by the BER vs. (1/AEVMS) results is captured by the AEVMS vs. $\Omega E_s$ results.

BER vs. (1/AEVMS) and AEVMS vs. $\Omega E_s$ plots can also be used to facilitate comparison of results obtained with different wireless testbeds. As we have mentioned, BER and AEVMS can be directly measured. Hence, BER vs (1/AEVMS) results obtained with different wireless testbeds can be directly compared without need for calibration between testbeds. Also, AEVMS vs. $\Omega E_s$ results can be used to compare how good the testbed is: for a given $\Omega E_s$ the best testbed is the one that has the lowest AEVMS. Metrics like BER, AEVMS, and $\Omega E_s$ which facilitate comparison between results obtained with different wireless testbeds and between experimental and simulation results are of great value for benchmarking and debugging.

## V. CONCLUSION

We presented a comprehensive study of beamforming MISO systems. We presented simulation, analytical, and experimental results, and analyzed implementation requirements and effect of channel and implementation-induced imperfections. Our results show that, using EGB codebooks, it is feasible to efficiently implement codebook-based beamforming MISO systems that have good performance. We also showed that the Mixed Codebook scheme simplifies the implementation complexity of WiMAX beamforming systems. Finally, we showed that the AEVMS is a relevant metric for performance analysis of beamforming MISO systems which facilitates comparison between theoretical and experimental results and can also facilitate comparison between experimental results obtained with different wireless testbeds.

## Appendix

We show how to obtain equations (11) and (12). To obtain (11) we rewrite

$$E\left[\frac{\left|\mathbf{v}[k-D]\mathbf{w}_b^{\dagger}\right|^2}{\left|\widehat{\mathbf{h}}[k-D]\mathbf{w}_b^{\dagger}\right|^2}\right] = E\left[\frac{|v_1[k-D]w_{b,1} + ... + v_T[k-D]w_{b,T}|^2}{\left|\widehat{\mathbf{h}}[k-D]\mathbf{w}_b^{\dagger}\right|^2}\right], \quad (24)$$

where $w_{b,i}$ and $v_i[k-D]$ denote the $i$-th entry of vector $\mathbf{w}_b$ and $\mathbf{v}[k-D]$ respectively. Since $v_i[k-D]$ and $w_{b,i}$ are independent and $E[v_i[k-D]] = 0$, crossterms in the expectation in (24) cancel. Then, using the fact that $E[|v_i[k-D]|^2] = 1$ and $||\mathbf{w}_b|| = 1$, (24) reduces to (11).

To obtain (12) we rewrite $\left|\widehat{\mathbf{h}}[k-D]\mathbf{w}_b^{\dagger}\right|^2 = ||\widehat{\mathbf{h}}[k-D]||^2 \max_{1 \le i \le N} \left|\widetilde{\mathbf{h}}[k-D]\mathbf{w}_i^{\dagger}\right|^2$, where $\widetilde{\mathbf{h}}[k-D] = \frac{\widehat{\mathbf{h}}[k-D]}{||\widehat{\mathbf{h}}[k-D]||}$. Since $\widetilde{\mathbf{h}}[k-D]$ and $\widehat{\mathbf{h}}[k-D]$ are independent [4], we can write

$$\begin{aligned} E\left[\frac{1}{\left|\widehat{\mathbf{h}}[k-D]\mathbf{w}_b^{\dagger}\right|^2}\right] &= E\left[\frac{1}{||\widehat{\mathbf{h}}[k-D]||^2}\right] E\left[\frac{1}{\max_{1 \le i \le N}\left|\widetilde{\mathbf{h}}[k-D]\mathbf{w}_i^{\dagger}\right|^2}\right] \\ &= \frac{1}{\Lambda(T-1)} E\left[\frac{1}{\max_{1 \le i \le N}\left|\widetilde{\mathbf{h}}[k-D]\mathbf{w}_i^{\dagger}\right|^2}\right], \quad (25) \end{aligned}$$

where we have used the fact that $||\widehat{\mathbf{h}}||^2 \sim$ gamma$(T, \Lambda)$ hence $\frac{1}{||\widehat{\mathbf{h}}||^2} \sim$ inverse gamma$(T, \frac{1}{\Lambda})$, and $E\left[\frac{1}{||\widehat{\mathbf{h}}||^2}\right] = \frac{1}{\Lambda(T-1)}$. Using the relation between correlation and chordal distance [29] we have that $\max_{1 \le i \le N}\left|\widetilde{\mathbf{h}}[k-D]\mathbf{w}_i^{\dagger}\right|^2 = 1 - \min_{1 \le i \le N} d^2(\widetilde{\mathbf{h}}[k-D], \mathbf{w}_i)$, where $d^2(\widetilde{\mathbf{h}}[k-D], \mathbf{w}_i)$ is the chordal distance between $\widetilde{\mathbf{h}}[k-D]$ and $\mathbf{w}_i$. We rewrite the expectation in (25) as

$$E\left[\frac{1}{\max_{1 \le i \le N}\left|\widetilde{\mathbf{h}}[k-D]\mathbf{w}_i^{\dagger}\right|^2}\right] = 1 + E\left[\frac{\min_{1 \le i \le N} d^2(\widetilde{\mathbf{h}}[k-D], \mathbf{w}_i)}{1 - \min_{1 \le i \le N} d^2(\widetilde{\mathbf{h}}[k-D], \mathbf{w}_i)}\right]. \quad (26)$$

Denote $Z = \min_{1 \le i \le N} d^2(\widetilde{\mathbf{h}}[k-D], \mathbf{w}_i)$, an approximation to the pdf of $Z$ (assuming an optimum codebook designed based on the Grassmannian criterion [3]) was found in [29] and is equal to

$p_Z(z) = N(T-1)z^{T-2}$ for $0 \le z \le (1/N)^{1/(T-1)}$. The expectation in (26) can be computed as

$$
\begin{aligned}
E\left[\frac{\min_{1 \le i \le N} d^2(\widetilde{\mathbf{h}}[k-D], \mathbf{w}_i)}{1 - \min_{1 \le i \le N} d^2(\widetilde{\mathbf{h}}[k-D], \mathbf{w}_i)}\right] &= \int_0^{\left(\frac{1}{N}\right)^{\frac{1}{T-1}}} \frac{z}{1-z} N(T-1)z^{T-2}dz \\
&= \frac{T-1}{T} N^{\frac{-1}{T-1}} \, {}_2F_1\left(1, T, 1+T, \left(\frac{1}{N}\right)^{\frac{1}{T-1}}\right), \quad (27)
\end{aligned}
$$

where ${}_2F_1$ denotes the Gauss Hypergeometric function and the result of the integration was found in [36]. Using (27), (26), and (25), we obtain (12).

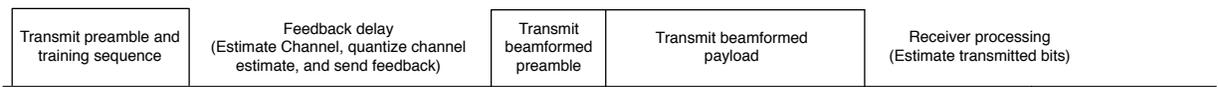

(a) Time diagram of a frame of a beamforming system which uses only one training sequence (1R system).

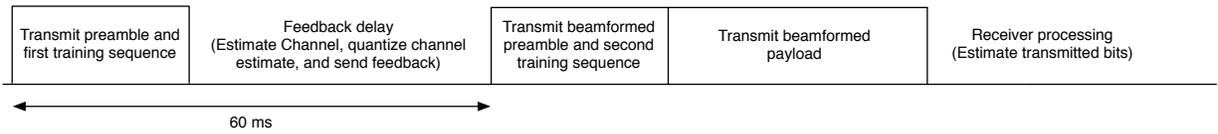

(b) Time diagram of a frame of a beamforming system which uses two training sequences (1.5R system).

Fig. 1. Time diagrams for different beamforming systems.

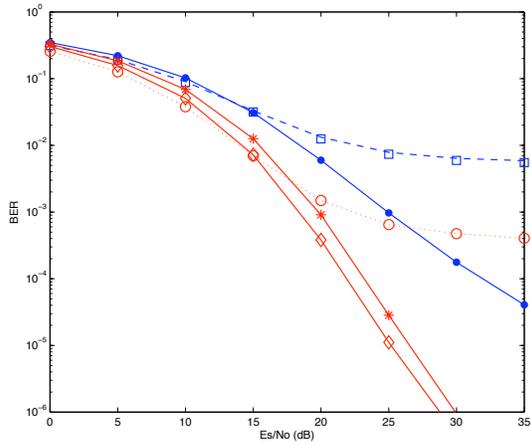

(a) BER vs. $E_s/N_o$

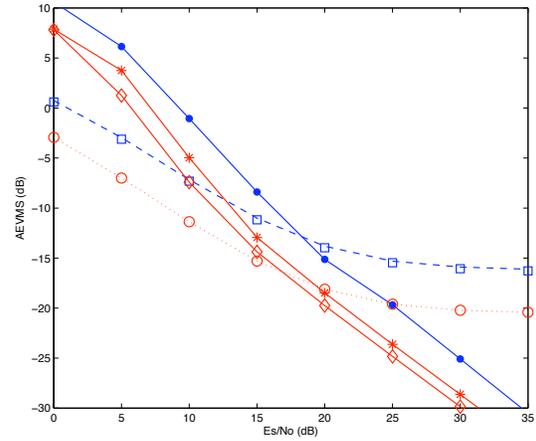

(b) AEVMS vs. $E_s/N_o$

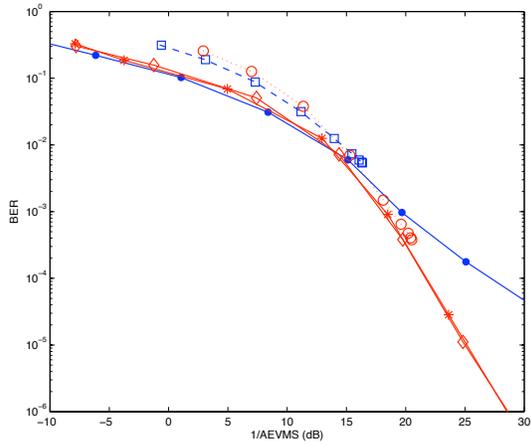

(c) BER vs. 1/AEVMS

(d) Legend for Fig. 2(a), Fig. 2(b), and Fig. 2(c).

- - - SN1 (Simulation): T = 2, N = 4, EGB Codebook, $\sigma_e^2 = (Es/No)^{-1}/2$, $\rho_d = 0.99$, 1R system.

□ SN1 (Analytical): T = 2, N = 4, Optimum Codebook, $\sigma_e^2 = (Es/No)^{-1}/2$, $\rho_d = 0.99$, 1R system.

—●— SN2 (Simulation): T = 2, N = 4, EGB Codebook, $\sigma_e^2 = (Es/No)^{-1}/2$, $\rho_d = 0.99$, 1.5R system.

○ SN3 (Simulation): T = 4, N = 64, EGB Codebook, $\sigma_e^2 = (Es/No)^{-1}/2$, $\rho_d = 0.99$, 1R system.

—★— SN4 (Simulation): T = 4, N = 64, EGB Codebook, $\sigma_e^2 = (Es/No)^{-1}/2$, $\rho_d = 0.99$, 1.5R system.

—◇— SN5 (Simulation): T = 4, N = ∞, $\sigma_e^2 = (Es/No)^{-1}/2$, $\rho_d = 0.99$, 1.5R system.

Fig. 2.   Simulation and analytical results showing the performance of a beamforming MISO system using BER and AEVMS for performance analysis. All results correspond to 16QAM modulation.

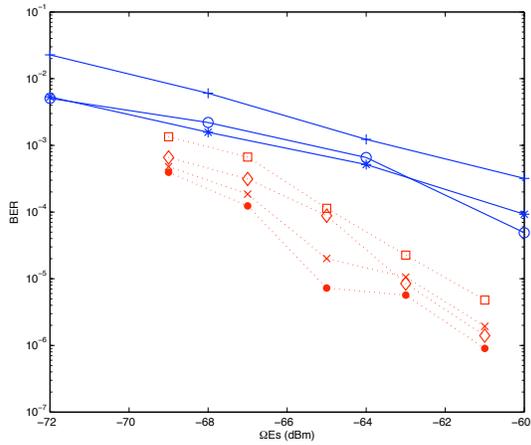

(a) BER vs. $\Omega E_s$

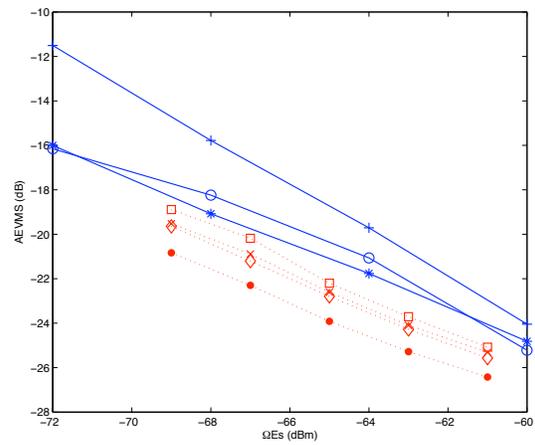

(b) AEVMS vs. $\Omega E_s$

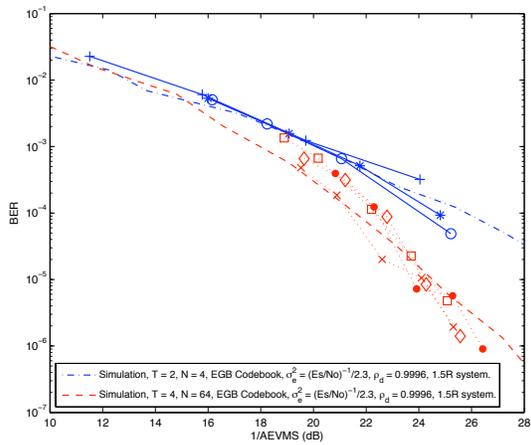

(c) BER vs. 1/AEVMS

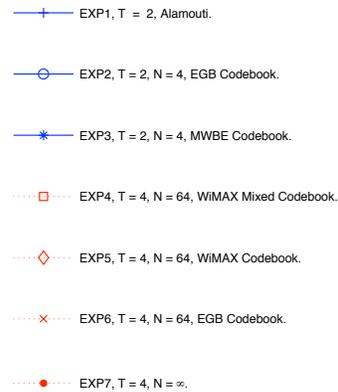

(d) Legend for Fig. 4(a), Fig. 4(b), and Fig. 4(c).

Fig. 3.   Emulator and simulation results showing the performance of a beamforming MISO system using BER and AEVMS for performance analysis. All results correspond to 16QAM modulation.

## TABLE I

CODEBOOK DESCRIPTION AND COMPARISON OF RESOURCES REQUIRED FOR CHANNEL QUANTIZATION FOR DIFFERENT CODEBOOKS.

| | Description | Resource Requirements | | |
|---|---|---|---|---|
| | | Resource | Total | Total for $N = 64$ $T = 4$. |
| MWBE Codebook | MWBE codebooks achieve the Welch bound on maximum cross-correlation between codewords defined in [26]. | Complex Mults. | $NT$ | 256 |
| | | Complex Adds. | $NT - N$ | 192 |
| | | Real Mults. | $2N$ | 128 |
| | | Real Adds. | $N$ | 64 |
| | | Negators | 0 | 0 |
| | | Mux 4 Inputs | 0 | 0 |
| | | Mux 9 Inputs | 0 | 0 |
| | | Relational | $N - 1$ | 63 |
| WiMAX Codebook | A codebook defined in the WiMAX standard [37]. | Complex Mults. | $NT$ | 256 |
| | | Complex Adds. | $NT - N$ | 192 |
| | | Real Mults. | $2N$ | 128 |
| | | Real Adds. | $N$ | 64 |
| | | Negators | 0 | 0 |
| | | Mux 4 Inputs | 0 | 0 |
| | | Mux 9 Inputs | 0 | 0 |
| | | Relational | $N - 1$ | 63 |
| EGB Codebook | We define an EGB codebook as a codebook that can be decomposed as $\mathbf{W} = \mathbf{GC}$, where $\mathbf{G}$ is an $N \times N$ diagonal matrix whose entires are real numbers and $\mathbf{C}$ is an $N \times T$ matrix whose entries belong to $\{1, -1, j, -j\}$. | Complex Mults. | 0 | 0 |
| | | Complex Adds. | $NT - N$ | 192 |
| | | Real Mults. | $2N$ | 128 |
| | | Real Adds. | $N$ | 64 |
| | | Negators | $2NT$ | 512 |
| | | Mux 4 Inputs | $2NT$ | 512 |
| | | Mux 9 Inputs | 0 | 0 |
| | | Relational | $N - 1$ | 63 |
| Tripolar Codebook | We define a Tripolar codebook as a codebook that can be decomposed as $\mathbf{W} = \mathbf{GC}$, where $\mathbf{G}$ is an $N \times N$ diagonal matrix whose entires are real numbers and $\mathbf{C}$ is an $N \times T$ matrix whose entries belong to $\{0, 1, j, -1, -j, 1+j, -1+j, -1-j, 1-j\}$. | Complex Mults. | 0 | 0 |
| | | Complex Adds. | $NT - N$ | 192 |
| | | Real Mults. | $2N + N$ | 192 |
| | | Real Adds. | $2NT + N$ | 576 |
| | | Negators | $4NT$ | 1024 |
| | | Mux 4 Inputs | 0 | 0 |
| | | Mux 9 Inputs | $2NT$ | 512 |
| | | Relational | $N - 1$ | 63 |

TABLE II
Experiment Conditions

| Parameter | Value |
|---|---|
| Number of transmitter antennas | $T = 2$ and $T = 4$ |
| Number of receiver antennas | 1 |
| Carrier frequency | 2.4 GHz |
| Number of subcarriers | 1 |
| Bandwidth | 625 kHz |
| ADC/DAC sampling frequency | 40 MHz |
| Pulse shaping filter | Squared Root Raised Cosine |
| SRRC roll-off factor | 1 |
| Symbol time | 3.2 $\mu s$ |
| Payload symbols per frame | 110 |
| Modulation | 16 QAM |
| Coding Rate | 1 (No error correction code) |
| Training signal energy per antenna | $E_p = E_s + 3.6$ dB |
| Feedback delay | $D = 60$ ms |
| Paths per emulated RF link | 3 |
| Model per path | Jake's model for all 3 paths |
| Fading Doppler per path | 0.1 Hz in all 3 paths |
| Delay per path | Path 1 = 0 $\mu s$ , Path 2 = 0.05 $\mu s$, Path 3 = 0.1 $\mu s$ |
| Relative path loss per path | Path 1 = 0 dB , Path 2 = 3.6 dB Path 3 = 7.2 dB |
| Delay correlation coefficient | For Jake's model is computed as $\rho_d = J_o(2\pi \cdot 0.1 \text{ Hz} \cdot 60 \text{ ms}) = 0.9996$ where $J_o(x)$ is the zeroth-order Bessel function of the first kind |